\documentclass[pss,fleqn]{w-art}
\usepackage{times}
\usepackage{w-thm}
\usepackage[]{graphicx}
\begin{document}
\renewcommand{\copyrightyear}{2006}
\DOIsuffix{theDOIsuffix}
\pagespan{1}{}


\title{Radiation hardness of diamond and silicon sensors compared\footnote{ Supported by EC Integrated Infrastructure Initiative Hadron Physics,
Project:RII3-CT-2004-506078}}


\author[W. de Boer]{Wim de Boer\footnote{Corresponding
     author: e-mail: {\sf wim.de.boer@cern.ch}, Phone: +49\,721\,608\,3593\, Fax:
     +49\,721\,608\,7930}\inst{1}}
\address[\inst{1}]{Universit\"at Karlsruhe, Physikhochhaus, Postfach 6980, D-76128 Karlsruhe, Germany} 
\author[J. Bol]{Johannes Bol\inst{1}}
\author[A. Furgeri]{Alex Furgeri\inst{1}}
\author[S. M\"uller]{Steffen M\"uller\inst{1}}
\author[C. Sander]{Christian Sander\inst{1}}
\author[E. Berderman]{Eleni Berdermann\inst{2}}
\author[M. Pomorski]{Michal Pomorski\inst{2}}
\author[M. Huhtinen]{Mika Huhtinen\inst{3}}
\address[\inst{2}]{Gesellschaft für Schwerionenforschung (GSI), Planckstr. 1, 64291 Darmstadt, Germany}
\address[\inst{3}]{CERN, CH-1211 Geneva 23}
\begin{abstract}
 The radiation hardness of silicon charged particle sensors is
compared with single crystal and polycrystalline diamond sensors, both
experimentally and theoretically. It is shown that  for Si- and C-sensors, the NIEL
hypothesis, which states that the signal loss is proportional to the
Non-Ionizing Energy Loss, is a good approximation to the
present data. At  incident proton and neutron energies well above 0.1 GeV the radiation damage
is dominated by the inelastic cross section, while at non-relativistic energies the
elastic cross section prevails. The smaller inelastic nucleon-Carbon cross section
and the light nuclear fragments imply that at high energies diamond is an order of
magnitude more radiation hard than silicon, while at energies below 0.1 GeV the
difference becomes significantly smaller.
\end{abstract}
\maketitle                   




\renewcommand{\leftmark}
{W. de Boer et al.: Radiation hardness of silicon and diamond sensors}

\section{Introduction}

The advent of relatively low priced CVD diamond wafers has opened up its use as
sensors for the detection of charged particles. The basic principle is the same as
in silicon sensors: metallic electrodes on  opposite sides are used to apply an
electric field, which drifts the electron hole pairs of an ionizing particle
towards the electrodes. The collected charge can be integrated by an integrating
amplifier or if fast signals are required, the ionization current can be directly
multiplied by a current amplifier. In harsh radiation environments, as expected
e.g. at future high intensity colliders, the radiation damage by hadrons or heavy
nuclei will be severe and silicon detectors are known not to survive too long.
Several collaborations have investigated the alternatives, either by going to
diamond sensors \cite{rd42} or other materials \cite{rd50} or cooling the silicon
sensors \cite{rd39}. The diamond sensors were shown to be rather radiation hard in
high energy proton beams, albeit with a signal level an order of magnitude below
the signals in silicon sensors. This improves by a factor around 3 with the advent
of single crystal (sCVD) diamond wafers, which become available now.

The radiation damage causes two effects: the change of dark current and the signal
decrease with increasing fluence of the detected particles. For silicon sensors the
strong increase of dark current requires a cooling of the detectors in order to
avoid reverse annealing and thermal runaway, while in diamond sensor the leakage
current even at room temperature is negligible and usually decreases after
irradiation. The signal decrease in silicon has been studied at various particle
energies and fluences and was found to be in most cases proportional to the
Non-Ionizing Energy Loss (NIEL) damage cross section, which is closely related to
the creation of lattice defects. At low beam energies $E$ the NIEL cross section is
dominated by the long-range Rutherford scattering, which falls like $1/E^2$ and
creates many small scale lattice displacements. At intermediate energies (above a
few MeV) the anomalous elastic Rutherford scattering from the nuclear interactions
between the incoming beam and the nuclei in the sensor starts to play a role, while
at  energies above  a few hundred MeV the inelastic cross section, which is almost
energy independent, dominates. The inelastic collisions fragment the nuclei and the
slow moving nuclear fragments lead to strong lattice defects by the Rutherford
scattering again.
 Impurities like oxygen, can reduce the signal losses by forming stable
non-trapping defects with the vacancies \cite{rose}, thus leading to a deviation
from the NIEL scaling hypothesis, which states that the radiation damage is
proportional to the NIEL damage cross section.

In diamond the expected increase in radiation damage by Rutherford scattering at
low energies has never been measured. This study was triggered by the observation
that the ionization signal in diamond sensors decreases surprisingly fast after
irradiation with 26 MeV protons.
 In this paper  the NIEL cross section in
diamond has been calculated and compared  with new data for the radiation damage
from protons and neutrons
 at low energies for the first time.

\section{Calculation of radiation damage}

The total energy loss of an impinging particle on a detector is strongly energy
dependent and is divided into an ionizing part and a non-ionizing part. The
ionizing part is used for the detection and the non-ionizing energy loss (NIEL)
represents the energy loss in phonons and lattice defects. The fraction of NIEL is
defined by the Lindhard partition function, which gives the ratio of the ionizing
energy loss over the total energy loss. This function is itself  strongly energy
dependent, as will be discussed below.

 The damage cross section $D$ is either given in $[MeVmb]$ or in
$[keVcm^2/g]$ and the corresponding KERMA (Kinetic Energy Release in Matter) is
given either by
\begin{equation}
  \label{eq1}
 KERMA(keV)=\Phi(\#/cm^2)\cdot wt(g)~\cdot~ D(keV cm^2/g)
 \end{equation} or
 \begin{equation}
  \label{eq2}
KERMA(MeV)=\Phi(\#/cm^2)~\cdot~(\#Atoms)~\cdot~10^{-27}(cm^2/mb)~\cdot~D(MeV mb).
\end{equation}
Here $\Phi$ is the flux of incident particles per $cm^2$ and $wt=\rho ~\cdot~
volume=\rho ~\cdot~ area ~\cdot~ target~ thickness$ is the weight of the target.
For silicon (diamond) $\rho=2.3 (3.5) g/cm^3$. The different units of the cross
sections are related by:
$(100 MeVmb)~\cdot ~ (10^3~keV/MeV)~\cdot~(10^{-27}
cm^2/mb)~\cdot~(mole/A~g)~\cdot~(6.023~\cdot~10^{23}/mole)= (60.216/A) keV~cm^2/g.$
 Here A is the relative atomic weight, which defines the weight of 1 Mol in $g/mol$, where 1 Mol contains
$6.023\cdot10^{23}$ atoms. For silicon (diamond)  A=28.086 (12).

According to the ASTM standard, the displacement damage cross section  for 1 MeV
neutrons is set as a normalizing value: $D_{n(1MeV)} = 95 MeVmb$ \cite{astm}. On
the basis of the NIEL scaling hypothesis the damage efficiency of any particle with
a given kinetic energy $E$ can then be described by the hardness factor k, defined
as
\begin{equation}k_{particle} (E)=D_{particle}/D_{n(1MeVn)}.\end{equation}
However, the normalizing value of $95 MeVmb$ is only valid for neutrons interacting
with silicon. For any other material 1 MeV neutrons will have another displacement
damage cross section. In order to compare the absolute cross sections between
silicon and diamond the NIEL values will be given in $MeVmb$ without normalization
to $1MeVn$.

\begin{figure}
\begin{minipage}{.5\textwidth}
\includegraphics[width=\textwidth,height=0.9\textwidth]{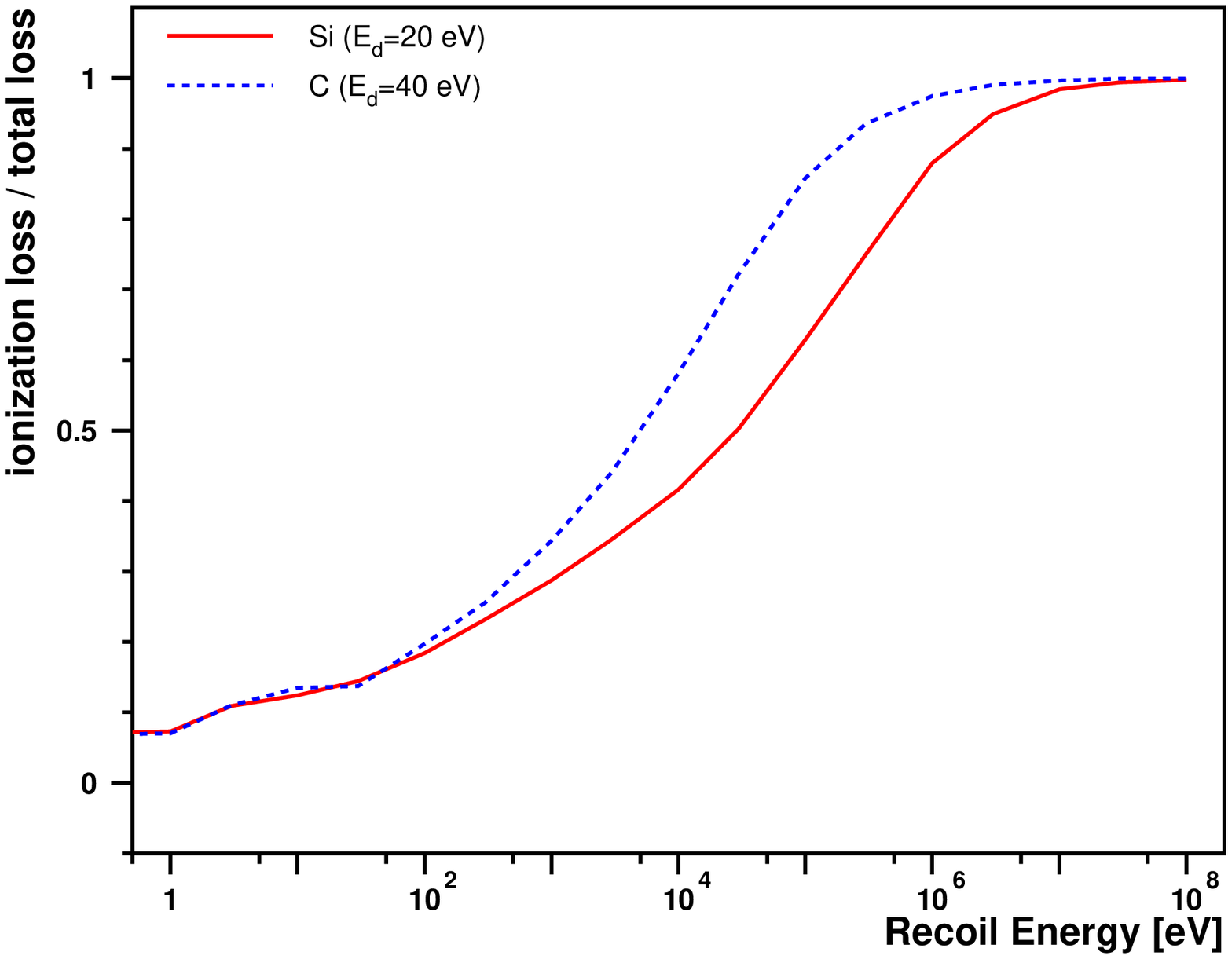}
\caption[]{{\it Lindhard partition function for silicon and diamond, as calculated
with the SRIM software package \cite{srim}. Notice that the NIEL is a significant
fraction of the total energy loss only for collision energies well below 10 MeV.}
\label{f1}}
\end{minipage}
\hfil
\begin{minipage}{.45\textwidth}
\begin{tabular}{|l|l|l|l|l|} \hline
$Z_{fr}$  &$Si_{fr}$ & $niel_{Si}$ & $C_{fr}$& $niel_C$ \\
\hline
 14 & 417 &4.2&0&0\\
 13 & 910 &9.1&0&0\\
 12 & 1384 &12.5&0&0\\
 11 & 1021 & 8.9&0&0\\
 10 & 1225 & 8.5&0&0\\
  9 &  265 & 1.4&0&0\\
  8 &  493 & 2.1&0&0\\
  7 &  398 & 1.3&0&0\\
  6 &  909 & 2.4&  698 & 0.8\\
  5 &  270 & 0.6&869&0.8\\
  4 &  383 & 0.7& 584&0.4\\
  3 &  662 & 0.7&1133&0.6\\
  2 &  11152& 4.4&10625&2.0\\
  1 &  46107&   0.9&30465&0.24\\
\hline
   Total& 65590 & 57.4&44374&4.8\\
 \hline
\end{tabular}
\captionof{table}{{\it A compilation of the  fragments, labeled by their charge
$Z$, as created by $10^4$ 10 GeV/c protons in Si and diamond sensors and the
contributions of these fragments to the NIEL cross section.} \label{t1}}
\end{minipage}
\end{figure}

A convenient package for the calculation of the energy loss of ions was
developed by J. Ziegler and is available from the web \cite{srim}. The package is
called SRIM, which stands for the Stopping and Recoil of Ions in Matter. The
stopping is important, since most damage is not caused by the primary particles,
but by the  recoil and nuclear fragments from the first atom hit, i.e. the
 primary knock-on atom (PKA). In our case the primary energies are so large
 that the PKA can create further defects, so a cascade develops. For inelastic collisions
 the nuclear fragments are so damaging that a cluster of defects develops.
 At low energies the elastic collisions dominate and only point like defects are created.

The energy loss  is determined by   collisions between two
colliding atoms, which requires the knowledge of the wave functions. The quantum
mechanical treatment  of such collisions is complicated and has to take the screening
of the nuclear charge by the electronic shells into account. Such calculations are
of utmost importance for e.g. the implantations needed for electronic circuits.
Therefore it is not
surprising that SRIM was developed at the IBM Research Laboratory. It has been studied
in great detail for many different nuclei and SRIM incorporates the
state-of-the-art calculation of the energy losses in matter, including the creation
of defects, interstitials, vacancies etc. Unfortunately the primary particles are considered
to have only elastic Coulomb interactions with the atoms of the target.

This is not enough for our purpose, since the impinging particles are usually
relativistic, which leads in addition to pure Coulomb scattering also to elastic
and inelastic scattering by nuclear interactions. However, the interaction of the
PKA and its nuclear fragments can be treated by the SRIM package. Therefore one can
calculate the energy spectrum of the nuclear fragments and nuclear recoils from
energetic collisions  and send these fragments and recoils as input particles to
the SRIM package, which then calculates  the NIEL contribution.  In such a way the
NIEL has been studied in detail for silicon by \cite{huhtinen}. His study was
adopted by adjusting the nuclear cross sections from silicon to diamond.

 Before showing the results a few points are in order.
The elastic Rutherford cross sections diverge for small scattering angles, so a cut
has to be imposed for the minimum momentum transfer. This cut was chosen in such a
way to effectively reproduce the NIEL at 1.3 MeV incident proton energy, as
calculated by SRIM. SRIM includes all the charge screening, which is not taken into
account by the Rutherford scattering formulae, so this cut is an effective way to
reduce the cross section and it has to be different for carbon and silicon.

For energy transfers below the displacement energy the energy loss is into phonons,
which is correctly taken into account in SRIM, but does not lead to defects. This
fraction of NIEL into phonons is somewhat energy dependent, so the defect
concentration is not strictly proportional to the NIEL cross section anymore.
Therefore deviations from the NIEL scaling hypothesis can be expected, but these
effects are estimated to be less than 20\% for the energy range considered. The
experimental uncertainties are considerably larger, so this effect is neglected.

The total NIEL can be calculated from the total energy loss via
the Lindhard function. The Lindhard functions for diamond and silicon are compared
in Fig. \ref{f1}, as calculated with SRIM. The averaged energy to create a lattice
displacement is about 15-20 eV for Si and in the range 37-47 eV for diamond
(depending on the direction, \cite{displacement}). As the figure shows, for an
incident energy above 10 MeV practically all the energy loss goes into ionization
energy. However, the electrons have a high mobility, so this damage is repaired
quickly. Low energies particles are created by the nuclear fragments from inelastic
collisions or the nuclear recoils from either elastic or inelastic scattering. For
silicon with an atomic number of 28 many nuclear recoils can be created, which
cause a large amount of NIEL, as shown in Table \ref{t1} for an incident proton
with a momentum of 10 GeV/c. The NIEL in C is shown as well. Here most of the
secondary particles are the light He nuclei, which cause a relatively small amount
of NIEL. This is the basic reason why diamond is an order of magnitude more
radiation hard at high energies. But at energies well below 100 MeV the cross
sections are dominated by the Rutherford scattering for charged incident particles.
Here the radiation hardness of silicon and diamond is expected to be a factor
$(Z_{Si}^2\rho_{Si})/(Z_C^2\rho_C)=(14^2\cdot 2.3)/(6^2\cdot 3.5) =3.6$ different,
which is approximately the case, as will be shown in the next section.

\begin{vchfigure}
\begin{minipage}{.5\textwidth}
\includegraphics[width=0.57\textwidth,angle=270]{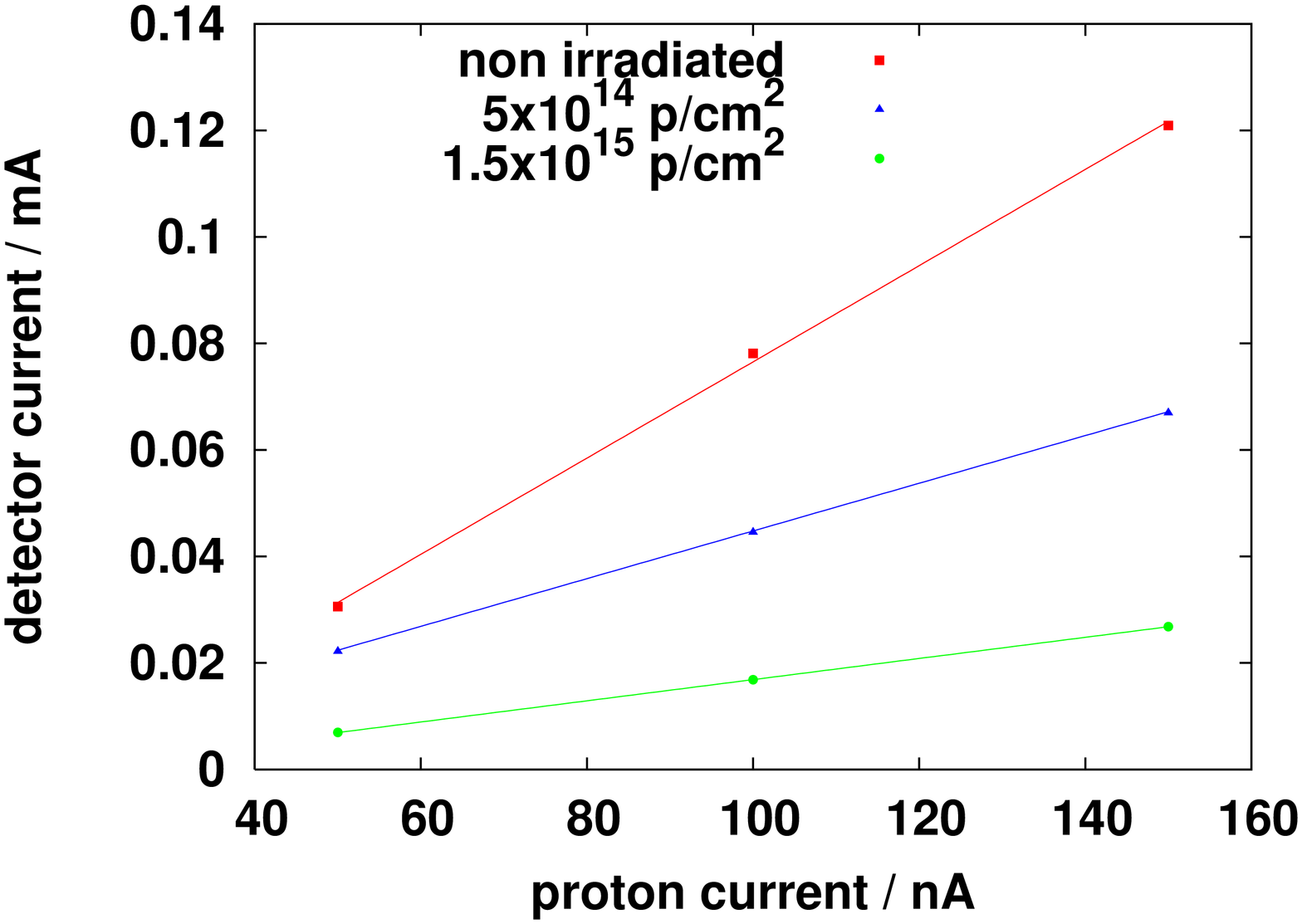}
\end{minipage}
\begin{minipage}{.5\textwidth}
\includegraphics[width=0.84\textwidth]{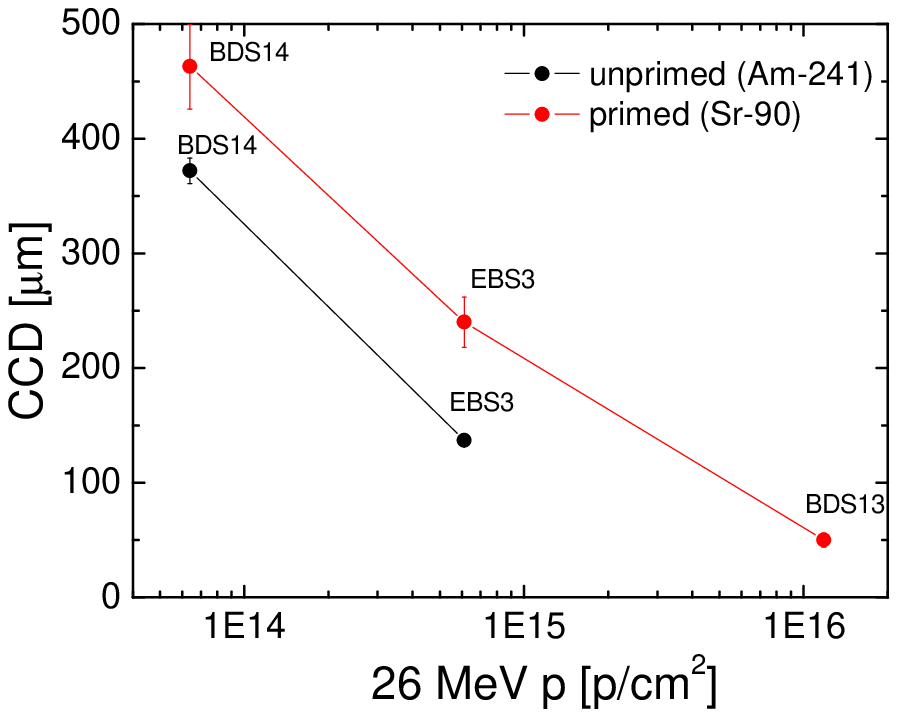}
\end{minipage}
\vchcaption{{\it Left: The observed ionization signal as function of the beam
current after three different fluences with 26 MeV protons. Right: the decrease of
the CCD for sCVD sensors as function of the fluence of 26 MeV $p/cm^2$. The CCD was
determined either from a Sr source with through going radiation from beta
particles, thus filling the traps (priming) or with an alpha source (Am), for which
the ionization is only created mainly in the first 10 $\mu$m of the sample.}
\label{f2}}
\end{vchfigure}
\begin{vchfigure}
\includegraphics[width=.32\textwidth,angle=270]{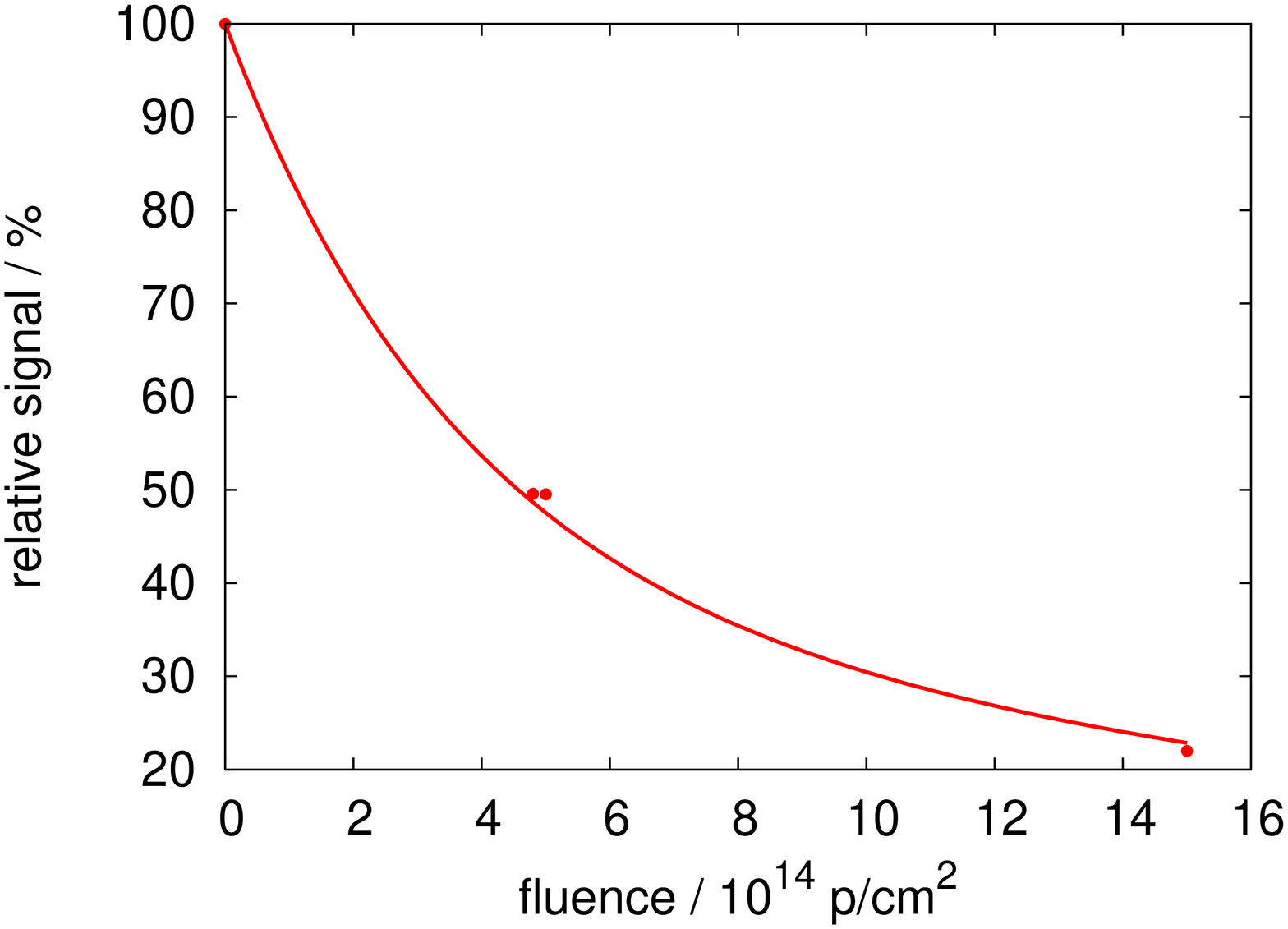}
\includegraphics[width=.32\textwidth,angle=270]{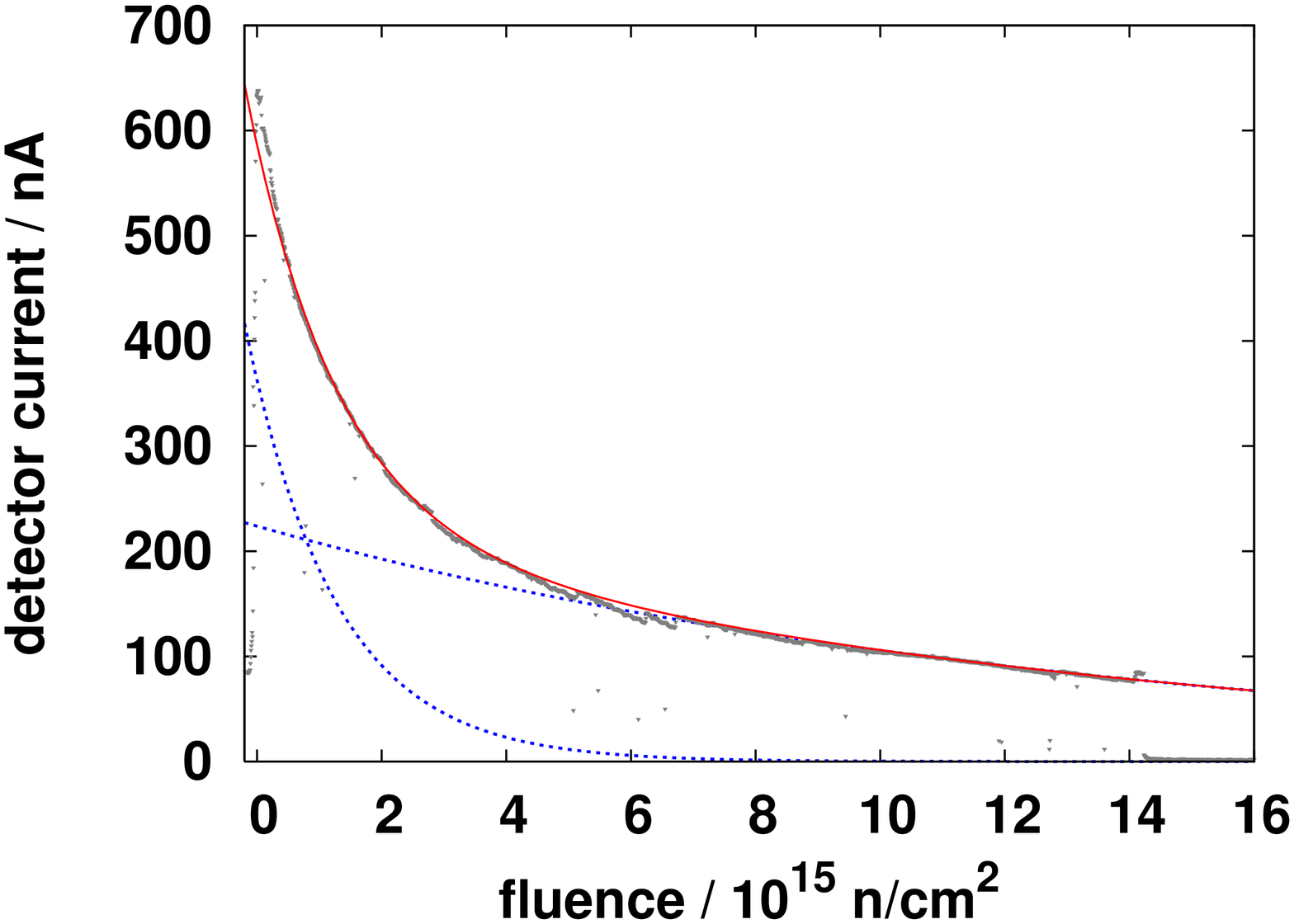}
\vchcaption{{\it The decrease of the ionization signal in a pCVD diamond sensor
after irradiation with 26 MeV protons (left) and 20 MeV neutrons (right).}
\label{f3}}
\end{vchfigure}

\begin{vchfigure}
\includegraphics[width=.45\textwidth,height=0.54\textwidth]{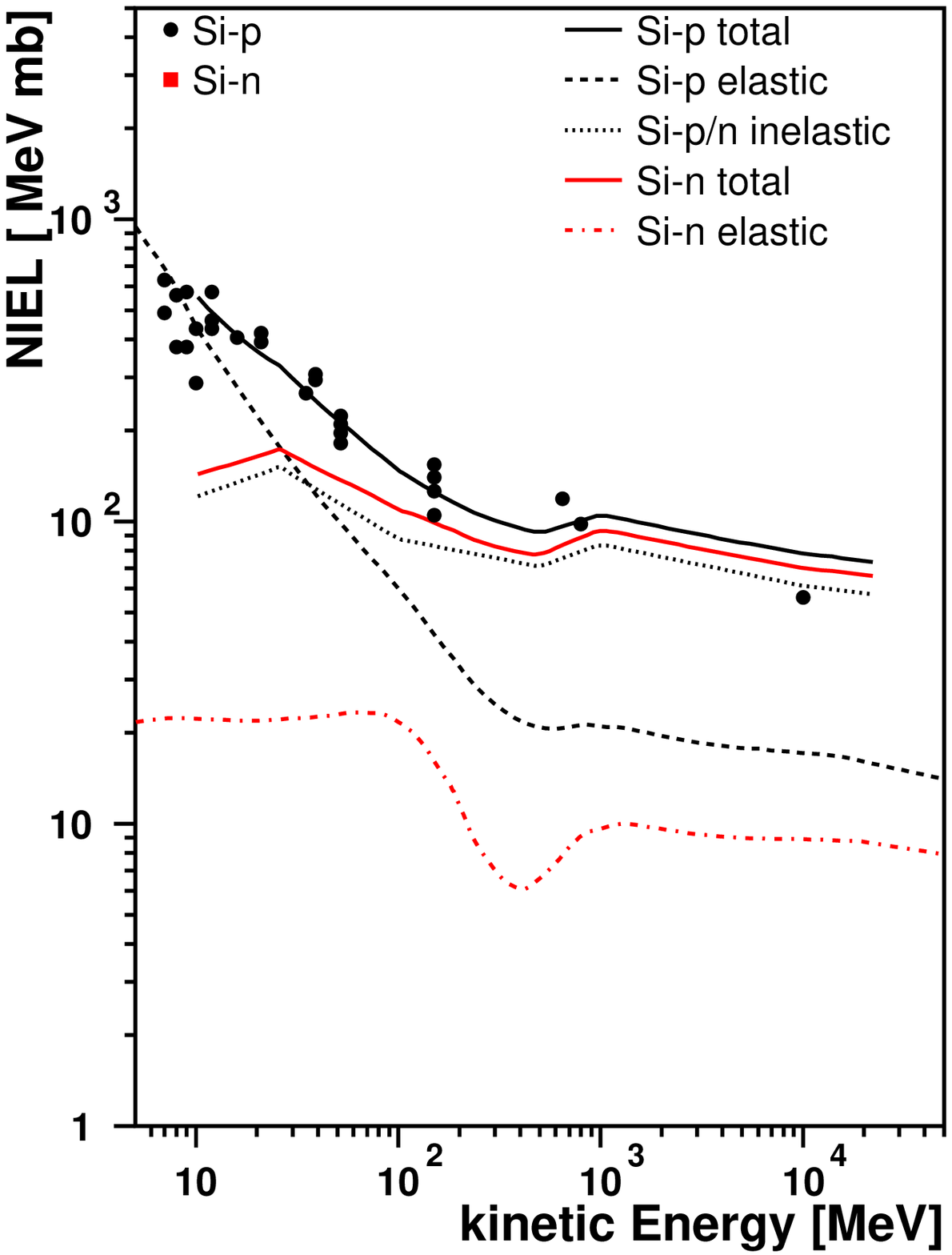}
\includegraphics[width=.45\textwidth,height=0.54\textwidth]{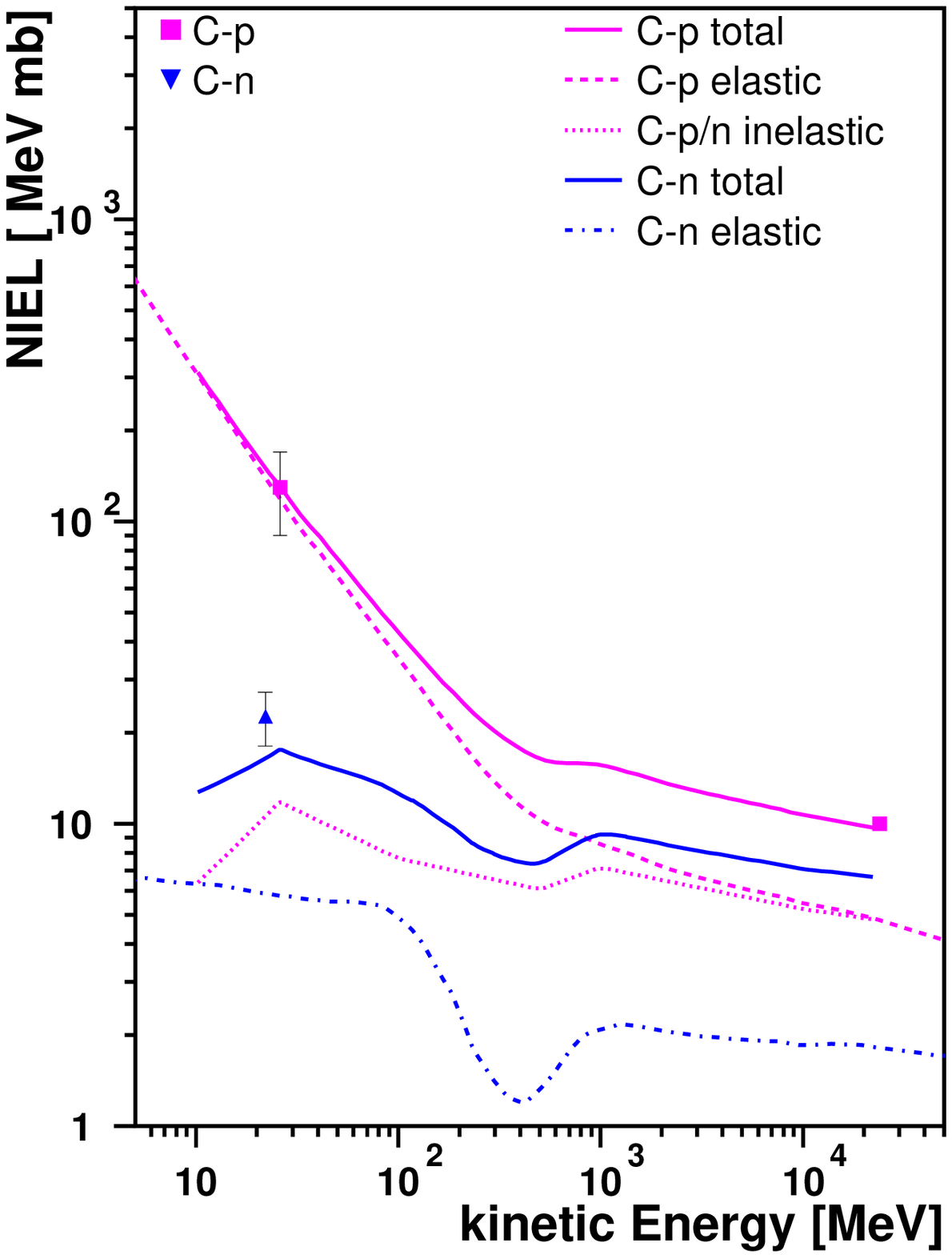}
\vchcaption{{\it NIEL damage cross section of Si (left) and Diamond (right) for
protons and neutrons (solid lines: upper one for p, lower one for n) as function of
the incident energy. The different cross section contributions from elastic and
inelastic scattering have been indicated as well.}  \label{f4}}
\end{vchfigure}

\section{Comparison with data}

The radiation hardness for diamond sensors has been measured by the RD42
collaboration using a $24~ GeV/c^2$ proton beam at CERN. They found that the
decrease in signal was around a factor 2 after a fluence of $(6\pm
2)\cdot~10^{15}~p/cm^2$ \cite{RD42_report}.  However, during a test of diamond
sensors with 26 MeV protons, the signal decreased much more rapidly, as shown  in
Fig. \ref{f2} (left). The curves were obtained by irradiating the sample with a
high intensity 26 MeV proton current and then reducing the current and measuring
the ionization current for different beam currents. After a fluence of $(4.5\pm
1.5)\cdot 10^{14}p / cm^2$ the ionization signal or the charge collection distance
(CCD) is reduced by a factor two.

There are two experimental worries concerning our method: a) is the online
measurement of the decrease of the ionization signal determining the real radiation
damage? b)for these studies we used rather cheap diamond sensors from \cite{E6},
which were not their sensor grade, but the "heat spreader" grade. Therefore the
results were repeated with a sensor grade pCVD diamond and a single crystal sCVD
diamond, all from \cite{E6}. The diamonds were either metallized with Al or (Au,Cr)
on C. The diamonds were characterized before and after irradiation using the
standard method of determining the charge collection distance with a radioactive
source. The reduction in signal for sCVD sensors follows a similar decrease as for
the online measurements with the pCVD sensors (see Fig. \ref{f2}, right), showing
that the bulk damage is an intrinsic property of the sensor material.

In order to check if this is due to the large increase in the cross section from
multiple Coulomb interactions the samples were irradiated with neutrons. Since the
neutrons yield much less ionization the ionization current could be monitored on
line without saturating the electronics. For the electronics a sensitive
"Current-to-Frequency" converter with a dynamic range from a few pA to a few mA was
user. This 8 channel module with an USB readout was developed by the LHC machine
group for the LHC beam monitoring with ionization chambers \cite{dehning}. The
irradiation was performed with the neutron beam at Louvain la Neuve with a mean
neutron energy of 20 MeV \cite{louvain}. The result is shown in Fig. \ref{f3}. A
twofold exponential is seen, indicating that a recombination effect is going on.
Details can be found in  \cite{mueller}. However, one observes that the signal
decrease by a factor two is reached after a fluence of about $(1.25\pm 0.25)\cdot
10^{15}$, which is already significantly better than the radiation hardness for
protons.

In order to check if these results make sense, the NIEL cross sections has been
calculated for diamond using the procedure outlined in the previous section. The
NIEL cross sections for Silicon (Diamond) are shown on the left (right) panel of
Fig. \ref{f4}. The silicon data were taken from \cite{huhtinen}, but scaled to fit
the present calculation at low energies. The present calculation changes somewhat
from the previous one because of using a newer SRIM version.
   The RD42 results at 24 GeV with $\Phi_{1/2}=(6\pm 2)\cdot 10^{15} p/cm^2$ were
used as normalization for the diamond results and the NIEL calculation was used to
determine the energy dependence. For the neutrons the normalization at 24 GeV was
taken to be 0.69 times $\Phi_{1/2}$ of protons, as expected from theory (see Fig.
\ref{f4}). Both, the large increase in cross section for the charged particles and
small increase for the neutrons, are  reproduced by the data.  Comparing the left
and right hand side shows that for low energies the difference in radiation damage
cross section between silicon and diamond is a factor of a few, while at high
energies the difference is an order of magnitude.

The agreement of the energy dependence of the NIEL cross section in comparison with
the data is remarkable, if one considers the large uncertainties in the
experimental data, e.g. from the priming of the diamond to fill the traps, as shown
in the right panel of Fig. \ref{f2}  or the non-exponential decrease of the signal,
as shown in Fig. \ref{f3}, which implies an ambiguity in the definition of the
radiation hardness. Note that only the shape of the energy dependence is relevant
here, since the data were scaled to fit the data at low energy for silicon and high
energy for diamond.

\section{Summary}
It is shown that  for Si- and C-sensors, the NIEL
hypothesis, which states that the radiation damage and the corresponding signal
loss is proportional to the Non-Energy-Energy-Loss (NIEL),
is a good approximation to the
present data. At incident proton and neutron energies well above
0.1 GeV the radiation damage
is dominated by the inelastic cross section, while at non-relativistic energies the
elastic cross section prevails. The smaller inelastic nucleon-Carbon cross section
and the light nuclear fragments imply that at high energies diamond is an order of
magnitude more radiation hard than silicon, while at energies below 0.1 GeV the
difference is significantly smaller. Such low energies are important even at high
energy hadron colliders, because the underlying events from the soft collisions of
spectator partons and secondary interactions in the detector material yield an
appreciable fraction of particles in this energy range.


\end{document}